\documentclass[aps,prd,twocolumn,a4paper,superscriptaddress,nofootinbib]{revtex4-1}
\usepackage{graphicx,caption,subcaption,amsmath,amsfonts,amssymb,multirow,extarrows,bm,acronym,float}
\usepackage[colorlinks,linkcolor=blue,citecolor=blue,urlcolor=blue ]{hyperref}
\usepackage[flushleft]{threeparttable}
\floatstyle{plaintop}
\restylefloat{table}
\newcommand{\PKU}{Kavli Institute for Astronomy and Astrophysics, Peking University, Beijing 100871, China}
\newcommand{\NAOCAS}{National Astronomical Observatories, Chinese Academy of Sciences, Beijing 100012, China}

\def\mnras{MNRAS}            
\def\prd{Phys.~Rev.~D}       
\def\prl{Phys.~Rev.~Lett}    

\begin{document}

\title{Comparison between time-domain and frequency-domain Bayesian inferences to inspiral-merger-ringdown gravitational-wave signals}

\author{Hai-Tian Wang}
\email[Corresponding author: ]{wanght@pku.edu.cn}
\affiliation{\PKU}
\author{Lijing Shao}
\email[Corresponding author: ]{lshao@pku.edu.cn}
\affiliation{\PKU}
\affiliation{\NAOCAS}

\date{\today}

\begin{abstract}
Time-domain (TD) Bayesian inference is important in ringdown analysis for
gravitational wave (GW) astronomy.  The validity of this method has been well
studied by \citet{2022arXiv220202941I}.  Using GW190521 as an example, we study
the TD method in detail by comparing it with the frequency-domain (FD) method as
a complement to previous study.  We argue that the autocovariance function (ACF)
should be calculated from the inverse fast Fourier transform of the power
spectral density (PSD), which is usually estimated by the Welch method.  In
addition, the total duration of the GW data that are used to estimate the PSD
and the slice duration of the truncated ACF should be long enough.  Only when
these conditions are fully satisfied can the TD method be considered
sufficiently equivalent to the FD method.
\end{abstract}

\maketitle

\acrodef{GW}{gravitational wave}
\acrodef{FFT}{fast Fourier transform}
\acrodef{LIGO}{Laser Interferometer Gravitational-Wave Observatory}
\acrodef{LVC}{LIGO-Virgo Collaboration}
\acrodef{BNS}{binary neutron star}
\acrodef{NS}{neutron star}
\acrodef{NR}{numerical relativity}
\acrodef{BH}{black hole}
\acrodef{BBH}{binary black hole}
\acrodef{NSBH}{neutron star-black hole}
\acrodef{IMBH}{intermediate mass black hole}
\acrodef{GR}{general relativity}
\acrodef{PN}{post-Newtonian}
\acrodef{SNR}{signal-to-noise ratio}
\acrodef{PSD}{power spectral density}
\acrodef{PDF}{probability density function}
\acrodef{ACF}{autocovariance function}
\acrodef{IMR}{inspiral-merger-ringdown}
\acrodef{QNMs}{quasinormal modes}
\acrodef{ISCO}{innermost-stable circular orbit}

\section{Introduction}\label{sec:intro}

Gravitational wave (GW) events detected by the LIGO-Virgo-KAGRA (LVK)
Collaboration \citep{LIGO_PRX2019,LIGO_O3a_PRX2020,2021arXiv211103606T} give us
an extraordinary opportunity to probe the gravitational physics in the strong
field.  The information of these GW events is hidden in the sea of noise
introduced by the detector itself, the surrounding environment, artificial
activities, and so on \citep{LIGOScientific:2019hgc}.  To extract information
from GW data, we usually perform Bayesian inference in the frequency domain
(FD).  Note that the raw data detected by the detectors are in the time domain
(TD).  Therefore, we need to perform a \ac{FFT} on the raw data for analysis in the
FD.  The \ac{FFT} requires GW data to be infinitely long or periodic on
the boundary, otherwise it may be affected by artifacts such as spectral
leakage.  For signals that do not satisfy these conditions, such as the ringdown
signal which has a sharp start, TD Bayesian inference has been developed
\citep{2021arXiv210705609I}.

The ringdown signal is the final stage of a GW signal and is emitted by the
oscillation of the remnant.  It is described by the \ac{QNMs} and decays very
quickly. It is usually expressed in the form of superimposed damped sinusoids
\citep{Schw_PRD_Vishveshwara1970,GW_APJL_Press1971,QNM_APJ_Teukolsky1973}.
Therefore, the ringdown signal is short enough to adopt the TD method, which
is, for long signals, usually limited by the computational cost of solving the
matrix inversion.  In parallel, there are some investigations which
developed different approaches to perform the FD Bayesian inference on the
ringdown signal \citep{2021PhRvD.104l3034F, 2022PhRvD.106d3005F,
2021PhRvD.103b4041B, Capano:2022zqm}.

Interestingly, \citet{2021PhRvL.127a1103I} and \citet{2022PhRvL.129k1102C} come
to  different conclusions, while there are only some technical differences in
the handling of the TD method.  The former team analyzed the ringdown signal of
GW150914 \citep{gw150914_PRL2016} with a sampling rate of $2048$ Hz and a slice duration of $0.5$ s, and they
found evidence for the first overtone mode with $3.6$-$\sigma$ confidence.  The
latter team analyzed the same ringdown signal, but with a different sampling
rate of $16$ kHz and a different slice duration of $0.1$ s, and they
demonstrated that the ``{\it claims of an overtone detection are noise
dominated}''.  In addition, there are some other technical details that can
affect the results of the TD method, such as the total duration of the GW data
used to estimate the \ac{PSD}.  It is challenging to conclude which setting is superior without knowing true values 
with TD methods alone.
Comparsions made with FD results on ringdown signals remain inconclusive as it is affected by the perviously mentioned sharp-edge effects in \ac{FFT}.

However, such a comparison is possible for the full inspiral-merger-ringdown
(IMR) signal of short duration, for example, in the GW190521 event
\citep{2020PhRvL.125j1102A}.  The duration of GW190521 is about $0.1$\,s, which
is sufficiently short for both FD and TD methods.  We therefore compare results
of the \ac{SNR} and the Bayesian inference for these two methods, varying 
settings with different technical details.  We find that the results of the TD
method agree well with those of the FD method only if some specific setting is used.

Apart from the studies mentioned above, the TD method has been widely used
to test the no-hair theorem \citep{2019PhRvL.123k1102I}, the black-hole area law \citep{2021PhRvL.127a1103I}, non-Kerr
parameters \citep{2021PhRvD.103l2002A, 2021arXiv211206861T, 2021PhRvD.104j4063W,
Cheung:2020dxo, Mishra:2021waw}, and black-hole thermodynamics
\citep{Hu:2021lbt}.  Although \citet{2021arXiv210705609I} have already studied
this method in detail, our study here can be seen as a complement.  It is
important for the ongoing LVK observing runs and future GW detectors such as
Einstein Telescope \citep{2010CQGra..27s4002P}, Cosmic Explorer
\citep{2019BAAS...51g..35R}, Laser Interferometer Space Antenna
\citep{LISA_arxiv2017}, TianQin \citep{TQ_2015,TianQin:2020hid}, and Taiji
\citep{1093nsrnwx116}.

This paper is organised as follows.  In Sec.~\ref{sec:noise} and Sec.~\ref{sec:comp}, we show
comparisons of noise and \ac{SNR} estimations, respectively.  In Sec.~\ref{sec:bayes}, we
show results of Bayesian inference from the TD method and the FD method.  In
Sec.~\ref{sec:conclusion}, we provide a brief summary and some discussion.
Throughout the paper, unless otherwise specified, we adopt geometric units where
$G=c=1$.

\section{Noise estimations}\label{sec:noise}

We use GW data provided by the Gravitational-Wave Open Science
Center\footnote{\url{https://gwosc.org/}} \citep{LIGO_O3a_PRX2020}.  There are
several options for the raw data, and we obtained the data with a duration of
$4096$ s at a sampling rate of $16$ kHz.  We can later truncate it to the length
that we need or resample it to a lower sampling rate.  
For the resampling algorithm, we use the one with the \texttt{Butterworth filter}, which is implemented in \texttt{PyCBC} (version $2.0.5$) \citep{2012PhRvD..85l2006A}.
In this study, we estimate the \ac{ACF} using different durations ($d_T$) and sampling frequencies ($f_{\rm s}$).

We assume that the noise around GW190521 is Gaussian and stationary after applying a high-pass filter at $f_{\rm low}=20$ Hz. 
Specifically, we use the \texttt{Finite Impulse Response} filter \citep{KhanFIR2020} with an order of $512$.
We then estimate the one-sided \ac{PSD} from downsampled data using the Welch method \citep{1967D.Welch} and the {\it inverse spectrum truncation} algorithm.
Note that all algorithms used above are implemented in the \texttt{PyCBC} package \citep{2012PhRvD..85l2006A}.

After the implementation of the high-pass filter, the noise can be well characterised by its \ac{PSD} or \ac{ACF}. 
There are three solutions to estimate the noise.
Firstly, in the FD method, the one-side \ac{PSD} ($S_n(f)$) is estimated with the Welch method \citep{1967D.Welch} and the {\it inverse spectrum truncation} algorithm, 
which are implemented in the \texttt{PyCBC} package \citep{2012PhRvD..85l2006A}.
In this case, the inner product of two signals $h_1(f)$ and $h_2(f)$ is
\begin{equation}\label{eq:inner_fd}
\langle h_1(f)|h_2(f)\rangle=4\int_{f_{\rm low}}^{f_{\rm s}/2}\frac{h^*_1(f)h_2(f)}{S_n(f)}{\rm d}f,
\end{equation}
where the asterisk in the upper right corner of $h_1(f)$ represents the complex conjugate of the signal.
The inner product is important since it determines the way to calculate \ac{SNR}s and likelihoods.

Then, there are two distinct TD methods, namely TTD1 and TTD2, employed in this study. 
Both methods utilize the same TD inner product, which is defined as 
\begin{equation}\label{eq:inner_td} 
\langle h_1(t)|h_2(t)\rangle=h_1(t)\mathcal{C}^{-1}h_2^{\intercal}(t). 
\end{equation} 
Herein, $\intercal$ located at the upper right corner of $h_2(t)$ gives the transpose of the signal, which is a discrete vector in real GW data analysis. 
The difference between these two methods arises from their respective computation approaches for the covariance matrix $\mathcal{C}$. 
Note that the covariance matrix is determined by the \ac{ACF}.

The \ac{ACF} of the TTD1 method is estimated from GW data directly,
\begin{equation}
\begin{aligned}\label{eq:ctt}
\mathcal{C}(k\Delta t)=&E[n(t)*n(t-k\Delta t)]\\
=&\frac{1}{K}\sum^{K-k}_{j=0} n(j\Delta t) n((j+k)\Delta t),
\end{aligned}
\end{equation}
where $K$ is the total number of data samples, $0\leq k\leq K-1$, $E[\cdot]$ denotes the expectation value, $n(t)$ represents discrete noise samples, and $\Delta t$ is the time intervals between samples. 
In this work, following \citet{2022PhRvL.129k1102C}, we calculate \ac{ACF} using the \texttt{get\_acf} function from \texttt{ringdown} (version $0.1$) package \citep{2021arXiv210705609I}.

The \ac{ACF} of the TTD2 method is estimated according to the Wiener-Khinchin theorem. 
One can calculate it from the one-side \ac{PSD}, $S_n(f)$, via
\begin{equation}\label{eq:ctf}
\begin{aligned}
\mathcal{C}(\tau)=&2\int^{\infty}_{f_{\rm low}}S_n(f)e^{2\pi if\tau} df\\
=&2\sum^{f_s/2}_{f_{\rm low}}S_n(f)e^{2\pi if\tau}\Delta f\, ,
\end{aligned}
\end{equation}
where $\tau=k\Delta t$.
Note that the \ac{PSD} is the same as that in the FD method.

As elucidated by \citet{2021arXiv210705609I}, the duration of \ac{ACF}s derived from Eq.~(\ref{eq:ctt}) and Eq.~(\ref{eq:ctf}) ought to exceed the intended analysis duration. 
Consequently, it becomes necessary to truncate these ACFs to our required duration. 
For instance, while the inner product employs a duration $d_t=0.5$ s, the estimated ACF's length from Eq.~(\ref{eq:ctf}) is $4$ s. 
We then proceed to truncate this $4$-s long ACF and retain only its initial segment with a span of $0.5$ s. 
This approach may yield results that deviate unexpectedly for both FD method and TTD2 method if not properly executed since the truncated ACF has less information than that of the original PSD.

For clarification, we summarize differences of these three methods as follows:
\begin{itemize}
\item \textbf{FD}: The \ac{PSD} is estimated with the Welch method. In this work, we use the one implemented in the \texttt{PyCBC} package. The definition of the inner product is shown in Eq.~\eqref{eq:inner_fd}.
\item \textbf{TTD1}: The \ac{ACF} is estimated according to Eq.~\eqref{eq:ctt}. In this work, we estimate it with the \texttt{get\_acf} function of the \texttt{ringdown} package. The definition of the inner product is shown in Eq.~\eqref{eq:inner_td}.
\item \textbf{TTD2}: The \ac{ACF} is estimated according to Eq.~\eqref{eq:ctf}. In this work, we use the Welch method and the {\it inverse spectrum truncation} algorithm implemented in the \texttt{PyCBC} package. The definition of the inner product is shown in Eq.~\eqref{eq:inner_td}.
\end{itemize}

\section{Comparisons}\label{sec:comp}

We have introduced the differences in noise estimation between the FD and TD
methods.  We then perform the Kolmogorov-Smirnov (KS) test on different methods
to see which one gives a better estimate of the noise.  We also show the results
of SNRs based on these noise estimates.

\subsection{KS tests}\label{subsec:ks}

To find an appropriate setting for the TD method, we perform the KS test for the
TD and FD methods with different settings.  The basic idea of the KS test is to
evaluate the credible level at which the whitened noise is consistent with a normal
distribution.  A KS $p$-value of $0.05$ means that we can exclude the
possibility that the tested distribution is not a normal distribution with
a $95\%$ confidence level.  The higher the $p$-value, the better its match with a
normal distribution.

We use two different durations, $d_T=64$ s and $d_T=4092$ s, and two different
sampling rates, $f_{\rm s} = 2048$ Hz and $f_{\rm s} = 16$ kHz.  Normally, we
split data into trunks of equal duration, for example, $4$ s.  Then we average
the \ac{ACF}s or the \ac{PSD}s that are estimated from these trunks.  We also
need to truncate the \ac{ACF} to the duration that we need, for example,
$d_t=0.5$ s.  This indicates that we assume that the duration of the GW signal
is enclosed within $d_t$.  

The stationary Gaussian noise $n(t)$ follows a multivariate normal distribution,
$n\sim \mathcal{N}\left(\mu,\mathcal{C}\right)$,
where $\mu$ is the mean value and $\mathcal{C}$ is the covariance matrix, which
is a Toeplitz matrix based on the \ac{ACF}.  We follow
\citet{2021arXiv210705609I} to Cholesky-decompose the covariance matrix into a
lower-triangular matrix $L$ and its transpose $L^{\intercal}$,
\begin{equation}\label{eq:cm}
\mathcal{C}=LL^{\intercal}.
\end{equation}
The whitened noise can be written as
\begin{equation}\label{eq:wn}
\bar n(t)=L^{-1}n(t).
\end{equation}

In the application of KS test, we partitioned the strain data into multiple segments of equivalent duration, specifically $4$ s. 
Consequently, for each method and setting, a set of segmented data was obtained. 
This allows us to compute an array of $p$-values and establish a distribution derived from the KS test.
We discard the truncated data that include the chirp signal, which means that we only
calculate the $p$-value for the off-source data.  If the \ac{ACF} describes the
noise well, $p$-values follow a rather flat distribution and most of
them should be greater than $0.05$.

\begin{figure}
\centering
\includegraphics[width=0.46\textwidth,height=6cm]{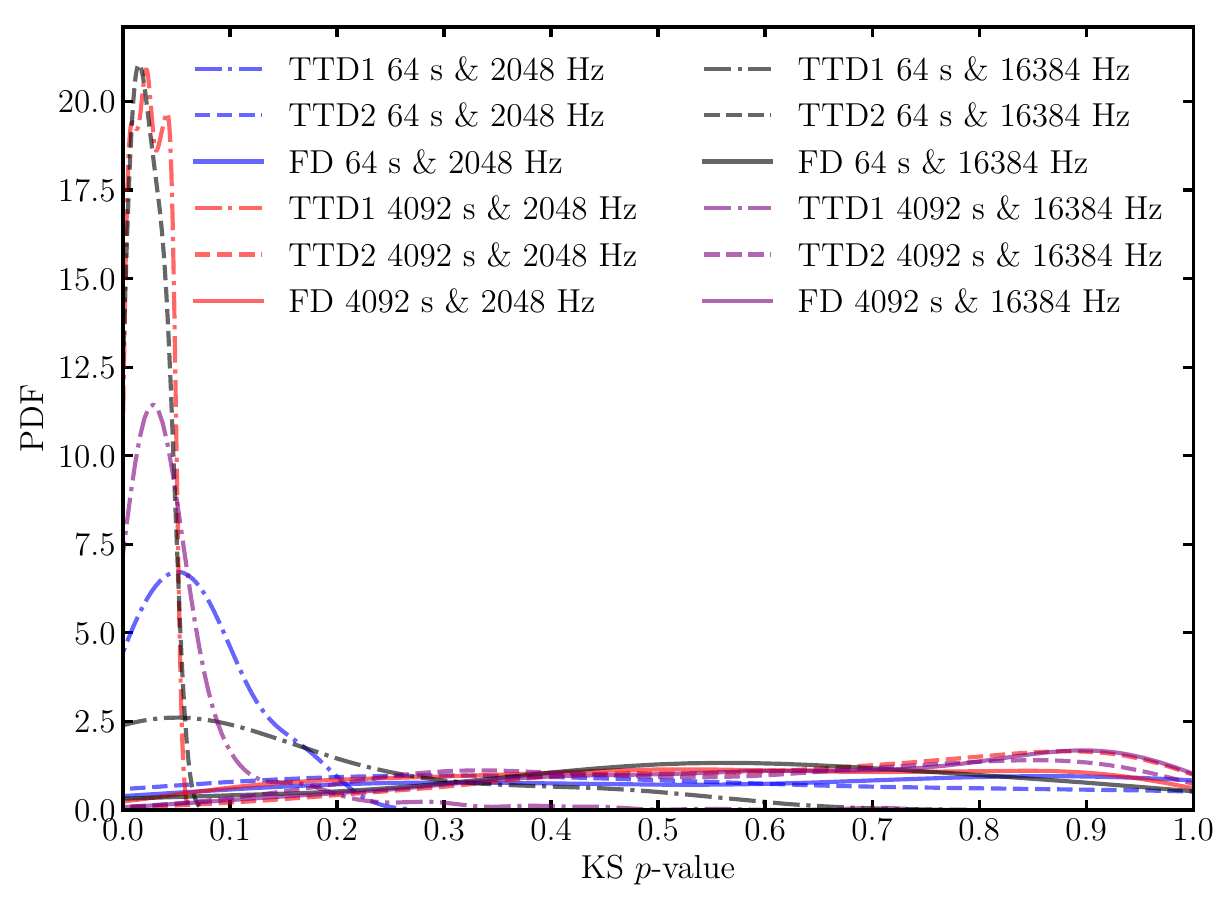}
\caption{
The distributions of the KS $p$-value for three methods with different
settings. For each method (same line style), we show the results of four
different settings in different colors.  The labels in the legend represent
the method used to obtain the \ac{ACF}, the total duration of the GW data
$d_T$, and the sampling rate $f_{\rm s}$.
}\label{fig:ksp}
\end{figure}

In Fig.~\ref{fig:ksp}, we show the effect of the total duration $d_T$ and the
sampling rate on different methods.  The durations of the truncated data are $d_t=0.5$ s and $d_t=4$ s for TD methods and the FD method, respectively.  
We can see that all the results of the FD method indicate
that the \ac{PSD}s describe the noise well.  However, for the results of the
TTD1 method, whose truncated \ac{ACF} is obtained from Eq.~(\ref{eq:ctt}),
almost all of them fail the KS test.  For the results of the TTD2 method, whose
truncated \ac{ACF} is obtained from Eq.~(\ref{eq:ctf}), only the case with a
duration of $64$ s and a sampling rate of $16$ kHz fails the KS
test. Compared with the results of Fig.~\ref{fig:snrs}, which will be
introduced later, it may seem strange that when the total duration is $d_T=64$ s
and the sampling rate is $16$ kHz, the results of the TTD2 method do not seem to
describe the noise well.  We find that this may be caused by the short total
duration, because 
the results become similar to those of the FD cases when we set
the total duration to $d_T=128$ s.  This highlights the importance of the
total duration $d_T$, as it affects the estimation of the \ac{ACF}.

\subsection{SNR estimations}\label{subsec:snr}

Another way to check the consistency among different settings is to compare the
\ac{SNR}s of a GW190521-like signal.  First, we generate the waveform in TD,
which can be written as $h(t,\lambda)$ where $t$ is the time and $\lambda$
represents the parameters of the waveform. The source parameters for the SNR
calculations are list in Table~\ref{tab:posts}.  We use the IMRPhenomXPHM
waveform model \citep{2021PhRvD.103j4056P} from \texttt{LALSuite}
\citep{lalsuite,swiglal}.  For TTD methods, it is easy to calculate the \ac{SNR}
based on Eq.~(\ref{eq:inner_td}) and Eq.~(\ref{eq:cm}),
\begin{equation}\label{eq:rho_td}
\rho^2=\left(L^{-1}h(t,\lambda)\right)^{\intercal}\left(L^{-1}h(t,\lambda)\right).
\end{equation}
For the FD method, we use \ac{FFT} to transform the TD waveform to FD and then
calculate the \ac{SNR} with
\begin{equation}\label{eq:rho_fd}
\rho^2=\langle h(f,\lambda)|h(f,\lambda)\rangle.
\end{equation}

\begin{figure*}
\centering
\begin{subfigure}[b]{0.48\linewidth}
\centering
\includegraphics[width=\textwidth,height=7cm]{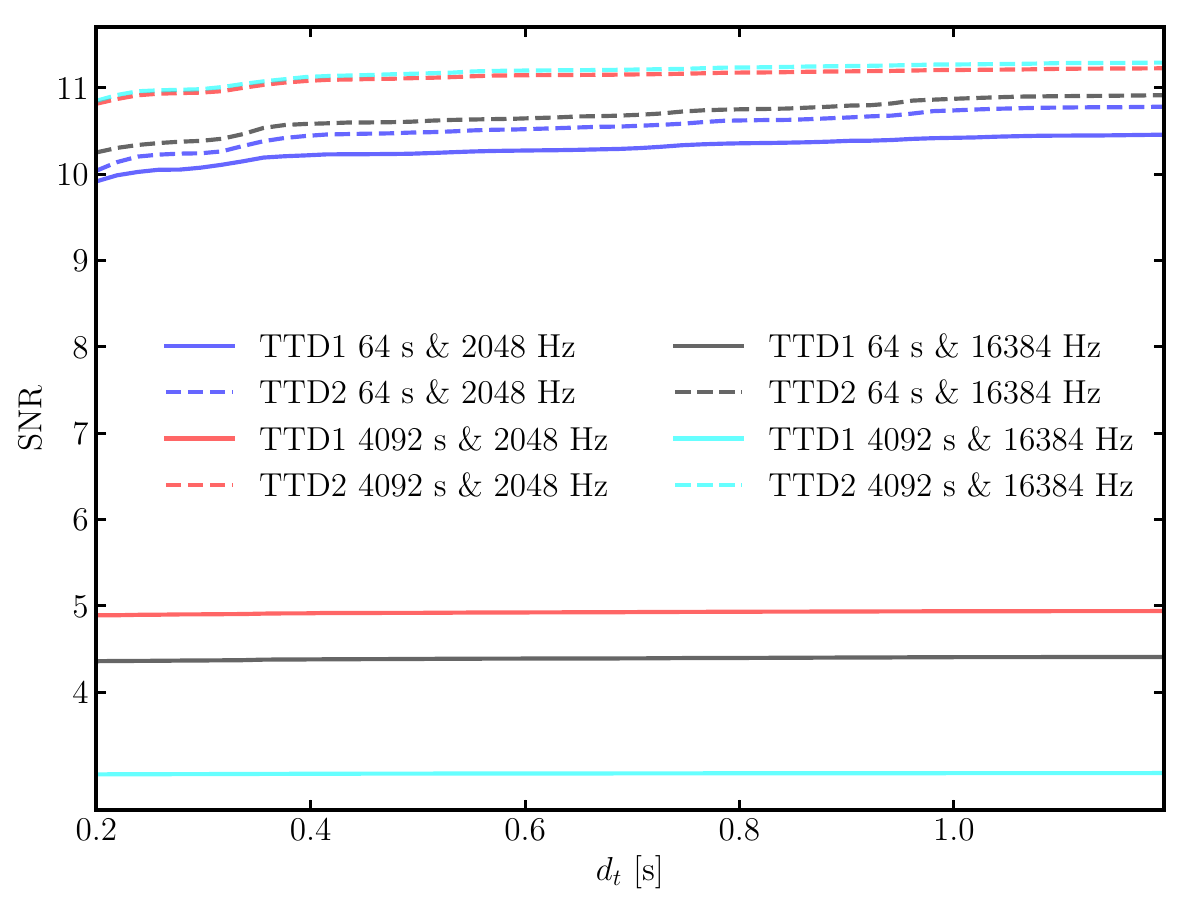}
\end{subfigure}%
\begin{subfigure}[b]{0.48\linewidth}
\centering
\includegraphics[width=\textwidth,height=7cm]{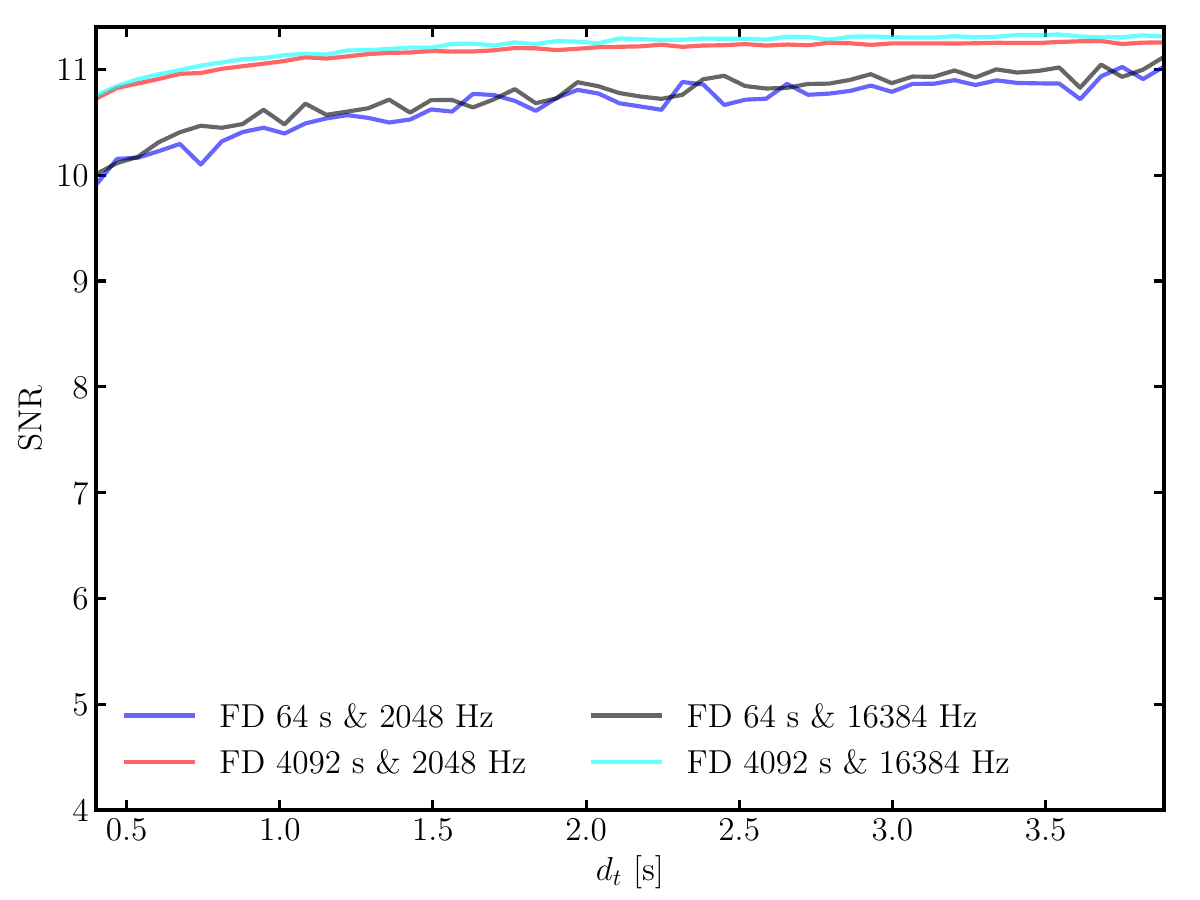}
\end{subfigure}%
\caption{
SNRs of a GW190521-like signal, varying with different durations of the truncated data $d_t$.
We show results based on ACFs from three different methods, with different total durations $d_T$ and different sampling rates $f_{\rm s}$.
}\label{fig:snrs}
\end{figure*}

As mentioned above, our GW190521-like event contains the full IMR signal,
although the inspiral of which is quite short.  So there is no abrupt start or
end for this signal and we expect different methods to give the same result.
When we calculate \ac{SNR}s using different methods, the trigger time of the
signal always locates in the centre of the data. Furthermore, we take the
results of the FD method as the true value, for this method is mature and
sufficiently verified. This is also supported by the results in
Fig.~\ref{fig:ksp} and Fig.~\ref{fig:snrs}.

Moreover we investigate the effect of the duration of the truncated data $d_t$ in
Fig.~\ref{fig:snrs}, where $d_t$ is also the duration of the signal generated by
the waveform model.  In Fig.~\ref{fig:snrs}, the TTD1 method provides large variance of \ac{SNR}s.  
The mean relative error of this method is approximately $73\%$, indicating potential inadequacies in the handling of \ac{ACF} estimation within the TTD1 method. In contrast, the Welch method exhibits greater reliability when applied to smaller data sets and demonstrates less susceptibility to nonstationarities.

From Fig.~\ref{fig:snrs}, we find that the sampling rate has little effect on the \ac{SNR}s of a
GW190521-like signal.  The mean relative errors between the \ac{SNR}s of the FD
method with different sampling rates are all smaller than $0.8\%$.  For the
\ac{SNR}s of the TTD2 method, the mean relative errors between different
sampling rates are smaller than $5\%$.  If we only change the total duration
$d_T$ for different methods, the mean relative errors of the \ac{SNR}s are
approximately $5\%$ for TTD2 and FD methods.  The results of TTD2 and
FD methods show that a larger $d_T$ gives a better estimate of the \ac{SNR}.
The \ac{SNR}s of the $64$-s long GW data are underestimated compared to the $4092$-s long cases.  It should be noted that the duration of the trancated data
$d_t$ also affects the results of different methods.  For these methods,
the relative error of \ac{SNR}s between the smallest $d_t$ and the largest
$d_t$ ranges from $4\%$ to $11\%$.

\section{Bayesian inference}\label{sec:bayes}

We now have a basic idea of how to use the TD method.  One of the main points is
that it should be performed with sufficiently long durations for both $d_T$ and
$d_t$.  Another important point is that the \ac{ACF} estimated by the TTD1 method may be biased in some improper setting.
To further investigate the consistency between two TD methods and the
FD method, we compare results of Bayesian inferences from these different
methods.  Specifically, we use a total duration of GW data $d_T=4092$ s
and a sampling rate $f_{\rm s} = 2048$ Hz for TD and FD methods.  
We select this particular configuration of duration and sampling rate due to its superior performance in both the KS test and SNR comparison, as detailed in Sec.~\ref{sec:comp}.
Furthermore, this choice aligns with the settings used in Ref.~\citep{Wang:2023mst}, demonstrating consistent results across different sampling rates.

The durations of the truncated data are $d_t=0.8$ s and $d_t=4$ s for the TD and FD methods
respectively.  Note that $d_t$ also represents the duration of the signal
generated by the waveform model.  On the one hand, the average relative error of
the \ac{SNR}s between the FD method and the TTD2 method is approximately
$0.3\%$, which is negligible.  Thus, we expect similar results from Bayesian inference using
these two methods.  On the other hand,
if one performs Bayesian inference with the biased \ac{ACF} given by the TTD1 method, we anticipate significantly different results.

In this section, we first introduce the Bayes theorem and some basic settings for Bayesian inference.
Then we show results based on more reasonable \ac{PSD}s, where the \ac{PSD} is estimated by the \texttt{PyCBC} package.

\subsection{Settings}\label{ssec:set}

To estimate the parameters $\lambda$ of GW190521 under a specific waveform model
$h(\lambda)$, we use the Bayesian inference based on the Bayes theorem,
\begin{equation}\label{eq:bt}
P(h(\lambda)|d,I)=\frac{P(d|h(\lambda),I)P(h(\lambda)|I)}{P(d|I)},
\end{equation}
where $P(h(\lambda)|d,I)$ is the posteriors of the parameters, $P(h(\lambda)|I)$
is their priors, $P(d|h(\lambda),I)$ is the likelihood, and $P(d|I)$ is the
evidence.  Note that the evidence of a specific model is a constant.  According
to \citet{Finn_1992}, the likelihood can be written as,
\begin{equation}\label{eq:lh}
\begin{aligned}
P(d|h(\lambda),I) =&P(d-h(\lambda)|I)\\
=&\exp\left[-\frac{1}{2}\langle d-h(\lambda)|d-h(\lambda)\rangle\right] \,.
\end{aligned}
\end{equation}
It is worth noting that, for the TD and FD methods, we calculate the likelihood according to Eq.~(\ref{eq:inner_td}) and Eq.~(\ref{eq:inner_fd}), respectively.  
For multiple detectors, one can simply multiply their likelihoods together.

We impose uniform priors on the redshifted component masses, and we sample from
the redshifted chirp mass and the mass ratio in ranges of $(10,200)\,M_{\odot}$
and $(0.04,1)$, respectively.  The prior on the luminosity distance is uniform
in the co-moving volume, with a range of $(50,10^4)$ Mpc.
For other parameters, we use the same priors as those used by the LVK
Collaboration~\citep{LIGO_O3a_PRX2020}.

\begin{figure*}
\centering
\includegraphics[width=0.96\textwidth,height=18cm]{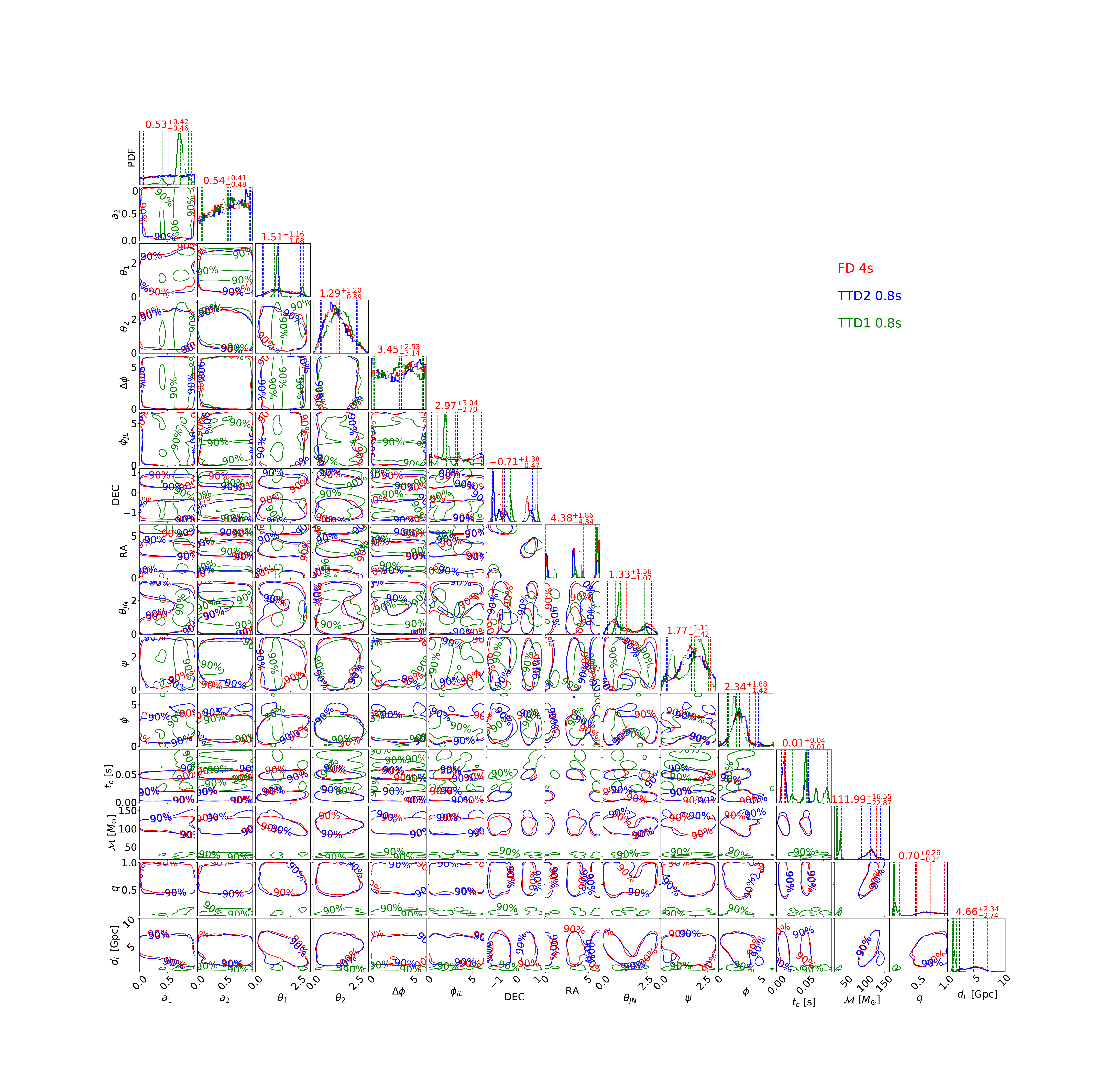}
\caption{
Posterior distributions for the parameters of GW190521 from three methods: the FD method (red), the TTD1 method (green), and the TTD2 method (blue). The durations for the FD method and two TD methods are $4$ s and $0.8$ s, respectively. The contours represent the $90\%$ credible level. The numbers on the diagonal are the median values and the $90\%$ credible measurements of each parameter in the FD method.
}\label{fig:posts}
\end{figure*}

We then implement the likelihood using the \texttt{Bilby} (version $1.2.1$) package
\citep{Ashton_APJ2019} and perform Bayesian inference using the \texttt{dynesty}
sampler \citep{Dynesty_MNRAS_Speagle2020} with $1000$ live points.  The minimum
(walks) and maximum numbers of sampling steps (maxmcmc) are $100$ and $5000$,
respectively.  

\subsection{Results}\label{ssec:res}

We perform the Bayesian inference using the FD method, the TTD1 method, and the TTD2 method.
Note that the downsampling algorithm and the high-pass algorithm used here are implemented in the \texttt{PyCBC} package.
We compare posteriors from these three methods in Fig.~\ref{fig:posts}. 
As expected, the FD method and the TTD2 method give
similar posterior distributions for almost all parameters.  Furthermore, the
log-Bayes factor between them is less than $0.1$, indicating that we can ignore
the differences.  We have also tried a duration of $1$ s for the truncated
\ac{ACF} of the TTD2 method and obtained similar results.  

However, for the TTD1 method, the results are quite different from those of the FD method and the TTD2
method.  The estimation of most parameters, such as the chirp-mass and the
luminosity distance, seems highly biased.  Furthermore, the log-Bayes factor
between the TTD1 method and the FD method is $\log_{10}\mathcal{B}^{\rm
TTD1}_{\rm FD}\approx -18$.  This means that the results from the TTD1 method
are very different from the FD and TTD2 methods.
This also agrees with what we expected that, a biased \ac{ACF} due to the improper setting of the TTD1 method leads to biased estimations.

\begin{table*}[hptb]
    \renewcommand{\arraystretch}{1.7}
\begin{ruledtabular}
\begin{tabular}{l | c | c c c | c c c | c}
 & FD & TTD2 & KLD & JSD & TTD1 & KLD & JSD & Injection\\ 
\hline
$a_1$ & $0.53^{+0.42}_{-0.46}$ & $0.53^{+0.42}_{-0.47}$ & $0.013$ & $0.003$ & $0.73^{+0.16}_{-0.33}$ & $1.387$ & $0.341$ & $0.70$\\ 
$a_2$ & $0.54^{+0.41}_{-0.48}$ & $0.59^{+0.36}_{-0.51}$ & $0.028$ & $0.007$ & $0.56^{+0.38}_{-0.47}$ & $0.024$ & $0.006$ & $0.96$\\ 
$\theta_1$ & $1.51^{+1.16}_{-1.08}$ & $1.32^{+1.28}_{-0.92}$ & $0.041$ & $0.010$ & $1.27^{+1.48}_{-0.18}$ & $2.327$ & $0.440$ & $1.07$\\ 
$\theta_2$ & $1.29^{+1.2}_{-0.89}$ & $1.22^{+1.25}_{-0.85}$ & $0.021$ & $0.005$ & $1.49^{+1.06}_{-1.02}$ & $0.067$ & $0.016$ & $1.54$\\ 
$\Delta\phi$ & $3.45^{+2.53}_{-3.14}$ & $3.43^{+2.59}_{-3.18}$ & $0.020$ & $0.005$ & $3.23^{+2.71}_{-2.87}$ & $0.029$ & $0.007$ & $0.52$\\ 
$\phi_{\rm JL}$ & $2.97^{+3.04}_{-2.7}$ & $3.24^{+2.82}_{-3.02}$ & $0.051$ & $0.013$ & $1.97^{+3.14}_{-1.1}$ & $0.950$ & $0.215$ & $3.47$\\ 
$\mathrm{DEC}$ & $-0.71^{+1.38}_{-0.47}$ & $-0.63^{+1.37}_{-0.58}$ & $0.288$ & $0.068$ & $-0.34^{+1.3}_{-0.8}$ & $1.731$ & $0.383$ & $0.68$\\ 
$\mathrm{RA}$ & $4.38^{+1.86}_{-4.34}$ & $3.32^{+2.92}_{-3.27}$ & $0.164$ & $0.043$ & $5.81^{+0.3}_{-4.71}$ & $1.817$ & $0.364$ & $3.51$\\ 
$\theta_{JN}$ & $1.33^{+1.56}_{-1.07}$ & $0.99^{+1.81}_{-0.74}$ & $0.084$ & $0.021$ & $0.99^{+1.42}_{-0.3}$ & $1.733$ & $0.321$ & $1.66$\\ 
$\psi$ & $1.77^{+1.11}_{-1.42}$ & $1.75^{+1.14}_{-1.45}$ & $0.022$ & $0.006$ & $1.91^{+0.83}_{-1.54}$ & $0.230$ & $0.056$ & $2.32$\\ 
$\phi$ & $2.34^{+1.88}_{-1.42}$ & $2.37^{+2.21}_{-1.42}$ & $0.029$ & $0.007$ & $2.03^{+1.53}_{-0.95}$ & $0.227$ & $0.052$ & $2.09$\\ 
$t_c$ [s] & ${t_0}^{+0.040}_{-0.006}$ & ${t_0}^{+0.041}_{-0.006}$ & $0.040$ & $0.010$ & $t_0+0.038^{+0.036}_{-0.025}$ & $0.987$ & $0.612$ & $t_0$\\ 
$\mathcal{M}$ [$M_{\odot}$] & $111.99^{+16.55}_{-22.87}$ & $113.24^{+26.21}_{-24.42}$ & $0.097$ & $0.029$ & $25.23^{+9.63}_{-3.05}$ & $0.0$ & $1.0$ & $126.6$\\ 
$q$ & $0.70^{+0.26}_{-0.24}$ & $0.68^{+0.28}_{-0.24}$ & $0.027$ & $0.007$ & $0.07^{+0.09}_{-0.02}$ & $0.015$ & $0.996$ & $0.89$\\ 
$d_L$ [Gpc] & $4.7^{+2.3}_{-2.7}$ & $4.4^{+2.5}_{-2.4}$ & $0.033$ & $0.008$ & $0.8^{+0.5}_{-0.2}$ & $0.755$ & $0.928$ & $2.75$\\ 
\end{tabular}
	\begin{tablenotes}
      \small
      \item *Note: $t_0=1242442967.413$ s is the trigger time of GW190521.
    \end{tablenotes}
\end{ruledtabular}
\caption{
The second column, the third column, and the sixth column show, respectively for
the FD method, the TTD2 method, and the TTD1 method, the medians and errors at
the $90\%$ credible level for the parameters of GW190521. To quantify the
differences between results of these three methods, we
calculate the KLD and JSD between their posterior samples against 
the FD method. The parameters in the ninth column are used to calculate SNRs as described in
Sec.~\ref{subsec:snr}. The definition of these parameters is the same as that
used in the \texttt{Bilby} package.
}\label{tab:posts}
\end{table*}

The parameter-estimation results are summarised in Table~\ref{tab:posts}.  To
quantify the differences between the results of those two TTD methods and the FD
method, we calculate the Kullback-Leibler divergence (KLD) and the
Jensen-Shannon divergence (JSD) between the posterior samples of them.  For two
posterior distributions $p$ and $q$, the KLD of them is defined by
\begin{equation}\label{eq:kld} 
D_{\rm KL}(p(x)|q(x))=\int p(x)\log_2\left[\frac{p(x)}{q(x)}\right]dx, 
\end{equation} 
and the JSD of them is defined by 
\begin{equation}\label{eq:jsd} 
D_{\rm JS}(p,q)=\frac{1}{2}\Big(D_{\rm KL}(p|s)+D_{\rm KL}(q|s)\Big), 
\end{equation}
where $s=(p+q)/2$.  As we can see in Table~\ref{tab:posts}, for the TTD2 method,
all JSD values are less than $0.07$, which means that it is acceptable to assume
that the difference between the results of these two methods is negligible.
However, for the TTD1 method, almost all JSD values are greater than $0.07$.
Thus, we conclude that, given current settings, the TTD1 method gives very
different results, compared to the other two methods.

\section{Discussion and Conclusion}\label{sec:conclusion}

In this paper, we investigate the performance of the TD Bayesian inference.  We
take the FD Bayesian inference as a comparison when analysing the GW190521
signal, which lasts for only $0.1$ s.  This event can be analysed by both the TD
and FD methods.  Note that such a comparison cannot be made with a ringdown
signal alone due to its abrupt start, nor can it currently be performed on a much longer
signal due to the computational cost of calculating the inverse of the
covariance matrix.  For example, for a $1$-s long signal, the dimension of the
covariance matrix will be as large as $16384\times 16384$ when the sampling rate
is $16384$ Hz.  The speed of calculating the likelihood is limited by the
computational cost, even if we compute the inverse of the covariance matrix
using the Cholesky method.

One of the key points of this investigation is how a truncated \ac{ACF} can
describe the noise as well as a circular \ac{ACF} does.  Therefore, we calculate
the KS $p$-values with two truncated \ac{ACF}s and one circular \ac{ACF}, which
are for the TTD method and the FD method, respectively.  We use them to
calculate \ac{SNR}s for a GW190521-like signal generated with the IMRPhenomXPHM
waveform model.  Compared to the results of the FD method, the results of the
TTD2 method are more reliable than those of the TTD1 method.  Specifically, the
largest relative error between the SNRs of the TTD1 and FD methods is
approximately $73\%$ with a non-optimal setting.  Furthermore, almost all
\ac{ACF}s for the TTD1 method fail in the KS test.  Note that one of the
\ac{ACF}s for the TTD2 method also fails in the KS test when the total duration
$d_T$ is not long enough.

The key factors in our comparison are the total duration $d_T$ for the noise
estimation, the duration of the truncated data $d_t$, and the sampling rate
$f_{\rm s}$.  For the total duration $d_T$, we find that $d_T=64$ s is not long
enough for the analysis of a GW190521-like signal.  The relative error caused by
this factor can be up to $5\%$ for the TTD2 method.  For the sampling rate, its
effect is negligible if the adopted total duration $d_T$ is large enough, i.e.
$d_T=4092$ s in our case.  For the duration of the truncated data $d_t$, the 
relative error can be as large as $11\%$.  From these analyses, it is
concluded that the TTD2 method should be performed with sufficiently long
durations for both $d_T$ and $d_t$, i.e., $d_t = 0.8$ s and $d_T = 4092$~s for a GW190521-like signal.

To further verify our results, we perform Bayesian inference using two TD
methods and the FD method.  For two TD methods (TTD1 method and TTD2 method), we
set $d_t=0.8\,{\rm s}$ and $d_T=4092\,{\rm s}$.  For the FD method, we set
$d_t=4\,{\rm s}$ and $d_T=4092\,{\rm s}$.  We confirm that, for the TTD2
method and the FD method, they have similar results, and the log-Bayes factor
between their posteriors is less than $0.1$.  The JSD values of each parameter
between these two methods are all less than $0.07$, meaning that the difference
between them is negligible.  

For the TTD1 method and the FD method, the
log-Bayes factor between them is about $-18$, and the JSD values of most
parameters are larger than $0.07$.  
However, we cannot conclude that the TTD1 method itself is not reliable.
It is posited that consistent outcomes can be achieved given the correct handling of the data. This assertion is substantiated by Fig.~\ref{fig:ksp} and Fig.~\ref{fig:snrs}, which illustrate results obtained from a total duration of $4092$~s at a sampling rate of $2048$ Hz, demonstrating consistency between the TTD1 method and the FD method. The reliability of the TTD1 method under different settings can further be verified through a KS test, as detailed in Sec.~\ref{subsec:ks}.
We do not delve deeply into this since the TTD2 method is both reliable and accessible.
We here aim to present results of the biased TTD1 method for comparison.
Overall, this analysis highlights the importance of performing the consistency check between different methods.

The TTD method is widely used in ringdown analysis.  Here, we show that it can
also be used in the full IMR analysis only when we handle it with extra care.
Furthermore, our analyses are helpful to understand the inconsistency
between the results of Ref.~\citep{2021PhRvL.127a1103I} and
Ref.~\citep{2022PhRvL.129k1102C}.
Based on our analyses, we got consistent results for ringdown analyses of GW150914 with different sampling frequencies, as shown in Ref.~\citep{Wang:2023mst}.
To allow for reproducibility, we have released codes for noise estimation at Ref.~\citep{acf_estimation}.

\begin{acknowledgments}
We thank Yi-Ming Hu for insightful discussions and the anonymous referee for constructive comments.
This work was supported by the China Postdoctoral Science Foundation
(2022TQ0011), the National Natural Science Foundation of China (12247152,
11975027, 11991053, 11721303), the National SKA Program of China
(2020SKA0120300), the Beijing Municipal Natural Science Foundation (1242018),
the Max Planck Partner Group Program funded by the Max Planck Society, and the
High-performance Computing Platform of Peking University.  HTW is supported by
the Opening Foundation of TianQin Research Center. 

This research has made use of data or software obtained from the Gravitational Wave Open Science Center (gwosc.org), a service of LIGO Laboratory, the LIGO Scientific Collaboration, the Virgo Collaboration, and KAGRA~\cite{KAGRA:2023pio}. 
LIGO Laboratory and Advanced LIGO are funded by the United States National Science Foundation (NSF) as well as the Science and Technology Facilities Council (STFC) of the United Kingdom, the Max-Planck-Society (MPS), and the State of Niedersachsen/Germany for support of the construction of Advanced LIGO and construction and operation of the GEO600 detector. 
Additional support for Advanced LIGO was provided by the Australian Research Council. 
Virgo is funded, through the European Gravitational Observatory (EGO), by the French Centre National de Recherche Scientifique (CNRS), the Italian Istituto Nazionale di Fisica Nucleare (INFN) and the Dutch Nikhef, with contributions by institutions from Belgium, Germany, Greece, Hungary, Ireland, Japan, Monaco, Poland, Portugal, Spain.
KAGRA is supported by Ministry of Education, Culture, Sports, Science and Technology (MEXT), Japan Society for the Promotion of Science (JSPS) in Japan; National Research Foundation (NRF) and Ministry of Science and ICT (MSIT) in Korea; Academia Sinica (AS) and National Science and Technology Council (NSTC) in Taiwan of China.
\end{acknowledgments}

\bibliographystyle{apsrev4-1}
\bibliography{imr}

\end{document}